\documentstyle[11pt,epsfig]{article} 
 
\newcommand{\bfvec}[1]{\hbox{\boldmath$#1$\unboldmath}} 
\begin{document} 
\title{MEDIUM MODIFICATION OF THE PION-PION INTERACTION 
AT FINITE DENSITY}
\author{D.  DAVESNE, Y.J. ZHANG and G. CHANFRAY \\ IPN Lyon, 43 Bd du 11
Novembre 1918, F-69622 Villeurbanne Cedex} 
\maketitle 
\begin{abstract} 
 
We discuss medium modifications of the unitarized pion-pion interaction in the
nuclear medium. We incorporate both the effects of chiral symmetry restoration
and the influence of collective nuclear pionic modes originating from the p-wave
coupling of the pion to delta-hole configurations. We show in particular that the
dropping of the sigma meson mass significantly enhances the low energy
structure created by the in-medium collective pionic modes.  
\end{abstract} 
 
 
\section{Introduction \hfill\ } 

Modifications of hadrons properties in nuclear and hot hadronic matter  
is one of the central subjects of present day nuclear physics. For instance,  
it has been suggested in \cite{Nor_88} that the in-medium pion-pion 
interaction might be significantly reshaped at density even below  
normal nuclear matter density. The basic mechanism is linked to the 
nuclear  collective pionic modes sometimes called pisobars originating 
from the p-wave coupling of the pion to delta-holes states. 
According to detailed calculation \cite{Gui_91,Uda_90} these collective 
modes are  needed  to explain charge exchange data \cite{Con_86,Hen_92},   
despite the peripheral character of these experiments.  This medium effect  
yields a softening of the pion dispersion relation  and consequently  
a modification of the two-pion propagator involved 
in the unitarized $T$ matrix describing pion-pion interaction at finite  
density. Indeed,  
on the basis  
of purely phenomenological models \cite{Cha_91}, an important reshaping      
of the pion-pion interaction in the scalar-isoscalar channel (sigma channel),  
producing a sizeable accumulation of strength near two-pion threshold, 
has been predicted. This problem has been reinvestigated with chiral symmetric  
models such as linear or non linear sigma models  \cite{Aou_94} with  
special emphasis on the  consistency between 
chiral symmetry constraints and unitarization. It has been soon realized  
that this medium effect is of considerable importance for the still open  
problem of nuclear saturation since an important part of the nucleon-nucleon  
interaction comes from correlated two-pion exchange  and several  
papers have brought extremely interesting results \cite{Dur_93,Rap_97,Mac_97}. 
A possible evidence for this reshaping of the $\pi-\pi$  
strength function is provided by the $\pi-2\pi$ data obtained  on various nuclei 
by the CHAOS collaboration 
at TRIUMF \cite{Bon_96}. Recent calculations show that the observed  
marked structure in the $\pi^+\pi^-$ invariant mass spectrum can be  
partially explained by this reshaping \cite{Sch_98,Kre_98}. These last results
have been  questioned in a recent paper \cite{Vic99} where  it is found
that pion absorption forces the reaction to occur at   lower peripheral
density.  .
 
However, what was   ignored in the previous approaches was the
possible medium  modification  of the basic $\pi-\pi$ interaction {\it i. e.}
the $\pi-\pi$ potential from   the underlying quark substructure or, in other
words, from the in-medium modification of hadronic properties associated to
chiral symmetry restoration.  
This may have considerable consequences since, applying by hand in the linear
sigma model pion-pion potential a Brown-Rho scaling of the sigma mass  yields a
significant enhancement of the near threshold structure in the $2\pi$ strength
function \cite{Aou_99}. To go further, it is important  to construct an
in-medium $\pi-\pi$ pion-pion potential This question has already been  
addressed \cite{Qua_95} in the framework of the Nambu-Jona-Lasinio (NJL) model
 at finite temperature. In this paper we will first use exactly the   same
scheme   but with a straightforward generalization at finite constituent  
quark density to be identified later with one third of the baryonic  ({\it
i.e.} nucleonic) density. In particular we will show that, using   a slightly
different prescription for the loop integrals,  the NJL   model regenerates
the linear sigma model pion-pion Born term amplitude. This result,   which is
numerically extremely close to the more involved calculational   scheme of ref.
\cite{Qua_95}, can be generalized at finite density   and/or temperature
provided the values of the pion mass, the sigma   mass and the pion decay
constant are replaced by their    in-medium values calculated in the NJL
model.    Using this scheme we are in position to study in a particularly
simple way  direct observable consequences  of both partial chiral symmetry 
restoration  such as dropping of the sigma meson mass and collective p-wave
pionic modes  by looking at the  in-medium pion-pion  interaction in the
scalar-isoscalar channel. This is of utmost importance since,  as emphasized
recently by Hatsuda {\it et al.}, the evolution of collective scalar-isoscalar
 modes,  {\it i.e.} the sigma meson, may reveal precursor effects   associated
to chiral symmetry restoration \cite {Hat_99}.  

From this   density dependent effective linear sigma model potential  
implemented with a phenomenological 
form factor, it is possible to construct a unitarized $\pi-\pi$  scalar-isoscalar  
amplitude which both preserves chiral symmetry constraints (Weinberg  
scattering length in the chiral limit) and reproduces experimental  
phase shifts. On top of precursor effects of chiral symmetry restoration, the 
inclusion of medium effects associated with the modification of 
the pion dispersion relation should be consistently done  
in the framework of the NJL  
model by direct coupling to constituent quarks populating    
the Fermi sea. However, the resulting pion p-wave polarizability calculated 
with quark-particles--quark-holes would completely miss the phenomenologically 
well established strong screening effects from short-range correlations 
($g'$ parameter). Incorporation of correlation effects in the  
 NJL model obviously requires a much more involved level of sophistication, 
hence loosing the simplicity which is one of its main interest.   
Furthermore we will calculate the  
in-medium two-pion propagator from standard pion-nucleus phenomenology. 
In other words, the p-wave  pion polarizability will be taken as its  
nuclear matter expression dominated in the  region of interest  
by the $\Delta$-hole piece corrected by screening effects. 
The underlying philosophy can be summarized in saying  
that the medium modified soft physics linked to chiral symmetry ($m_\pi, f_\pi$,  
low energy $\pi-\pi$ potential ) 
is calculated within the NJL model while p-wave physics yielding  
pionic nuclear collective modes is described through standard  
nuclear phenomenology.

\section{The $\pi - \pi$ Interaction from the NJL Model \hfill\ } 
To examine the $\pi - \pi$ interaction at finite density (and finite 
temperature), we start from the SU(2) version of the well-known NJL  
model\cite{Nam_61}: 
\begin{equation}                
{\cal L}= \bar{\psi}(i\not\!\partial-m_0)\psi 
         +g[(\bar{\psi}\psi)^2+(\bar{\psi}i\gamma_5\bfvec{\tau}\psi)^2] \ , 
\label{NJL}\end{equation} 
where $g$ is a coupling strength of dimension [Mass]$^{-2}$, and $m_0$ is 
the current quark mass. To maintain the asymptotic freedom at high energies, 
we simulate the situation by regarding the coupling constant in 
Eq. (\ref{NJL}) as  
a momentum dependent one g(p): 
\begin{equation}		
g(p)=g\prod_{i=1}^4\theta(\Lambda-|{\bf p}_i|) \ , 
\end{equation} 
where ${\bf p}_i$'s are the momenta of quarks and $\theta$ is the step function. 
Three parameters, namely $g$, $\Lambda$, and $m_0$ are determined to  
reproduce the pion mass $m_{\pi}$, the pion decay constant $f_{\pi}$, and  
the mass of sigma meson $m_{\sigma}$. Here for $m_{\pi}$ and $f_{\pi}$ we  
choose the well-accepted values, and for $m_{\sigma}$ we select two  
different empirical values to test the influence of the input data on our  
final results. In Tab. \ref{Paramt}, we list two sets of values for  
$m_{\pi}$, $f_{\pi}$ and $m_{\sigma}$, and also $g$, $\Lambda$ and  
$m_0$ which are fixed respectively to give these two sets of values of  
$m_{\pi}$, $f_{\pi}$ and $m_{\sigma}$. 
In the parameter fitting, we follow the formalism in the  
paper of Hatsuda {\it et al.} \cite{Hat_87}, where the three-momentum  
cut-off was used. With the lowest value of $m_\sigma=700\, MeV$, 
one has a set of parameters
very close to the one of Quack {\it et al.} \cite{Qua_95} used for finite temperature
studies. In the following mainly devoted to finite density, 
we will prefer the second set
with $m_\sigma=1\,GeV$  for two reasons. On one hand, this 
second set gives a higher  density for
chiral symmetry restoration (about $2.6\rho_0$ in place of $1.5\rho_0$)
and on the other hand, the values of $m_\sigma=1\, GeV$ has been 
successfully used in 
previous works to fit the vacuum $\pi-\pi$ data \cite{Aou_94,Sch_98}.
  
In Tab. \ref{Paramt}, we also include the results with  different schemes of  
calculation, the ``exact'' one and the ``Simplified'' one. For the ``exact''  
scheme, we just do the same calculation as Quack {\it et al.} \cite{Qua_95} 
but at finite density, while  
for the ``Simplified'' one, we neglect the $q^2$-dependence in all of the  
integrals (for details see Sec. 3) in our calculation,  
where $q$ is the external momentum of pion and sigma mesons. 
 
The main purpose of this section is to get the density-dependence on the  
$\pi$-$\pi$ scattering length, $a^I$, which is related to the $\pi$-$\pi$  
scattering amplitude ${\cal T}^I$ at threshold as: 
\begin{equation}		
a^I=-\frac{1}{32\pi m_{\pi}}{\rm Re}{\cal T}^I 
\end{equation} 
where $I$ is the total isospin.  
 
To the lowest order in $1/N_c$ ($N_c$, the number of  
quark colors), the invariant amplitude ${\cal T}_{ab;cd}$ of the  
$\pi - \pi$ scattering process ($a$, $b$ and $c$, $d$ are the isospin indices) 
is calculated from the box and $\sigma$-propagation diagrams shown in  
Fig. \ref{ScatFD} (see Fig. 1 in Ref. \cite{Qua_95}). Following the notation  
of Ref. \cite{Qua_95}, we have~:  
\begin{equation}		
{\cal T}_{ab;cd}=<cp_c;dp_d|{\cal T}|ap_a;bp_b> 
	       =A(s,t,u)\delta_{ab}\delta_{cd}+B(s,t,u)\delta_{ac}\delta_{bd} 
	        +C(s,t,u)\delta_{ad}\delta_{bc} \ , 
\end{equation}			 
where $s$, $t$ and $u$ are the usual Mandelstam variables: 
$s=(p_a+p_b)^2$, $t=(p_a-p_c)^2$ and $u=(p_a-p_d)^2$.  
 
Due to the crossing relations, there are three possibilities for the box 
diagram \cite{Sch_95} which, after a direct evaluation, gives: 
\begin{eqnarray}                
({\cal T}_1)_{ab;cd} 
    &=& -(\delta_{ab}\delta_{cd}+\delta_{ac}\delta_{bd}-\delta_{ad}\delta_{bc}) 
        [4N_cN_fig^4_{\pi qq}][I(0)+I(p)-p^2K(p)] \nonumber \\ 
({\cal T}_2)_{ab;cd} 
    &=& -(\delta_{ab}\delta_{cd}-\delta_{ac}\delta_{bd}+\delta_{ad}\delta_{bc}) 
        [4N_cN_fig^4_{\pi qq}][I(0)+I(p)-p^2K(p)] \\
({\cal T}_3)_{ab;cd} 
    &=& -(-\delta_{ab}\delta_{cd}+\delta_{ac}\delta_{bd}+\delta_{ad}\delta_{bc}) 
        [8N_cN_fig^4_{\pi qq}][I(0)+p^4L(p)/2 -2p^2K(p)] \nonumber
\end{eqnarray} 
in terms of integrals $I(0)$, $I(p)$, $K(p)$ and $L(p)$ (see appendix). 
 
The diagram with intermediate $\sigma$ propagation shown in Fig. \ref{ScatFD} 
can be simply expressed in terms of its components.  
By combination, one has~: 
\begin{eqnarray}		
({\cal T}_4)_{ab;cd} &=& - \delta_{ab}\delta_{cd}\,g^4_{\pi qq} 
                [\Gamma^{\sigma\pi\pi}(p,-p)]^2D_{\sigma}(2p) \nonumber \\ 
({\cal T}_5)_{ab;cd} &=& -(\delta_{ac}\delta_{bd}+\delta_{ad}\delta_{bc})g^4_{\pi qq} 
	        [\Gamma^{\sigma\pi\pi}(p,p)]^2D_{\sigma}(0) 
\end{eqnarray} 
where  
\begin{eqnarray}                
\Gamma^{\sigma\pi\pi}(p,-p) &=& -8N_cN_fmI(p) \nonumber \\ 
\Gamma^{\sigma\pi\pi}(p,p)  &=& -8N_cN_fm[I(0)-p^2K(p)] 
\end{eqnarray} 
are the $\sigma$-$\pi$-$\pi$ vertex in the $s$-channel and $t$-channel  
respectively, and 
\begin{equation}  		
D_{\sigma}(k)=\frac{i}{2N_cN_f[(k^2-4m^2)I(k)-m^2_{\pi}I(m_{\pi})]} \ . 
\end{equation} 
is the sigma meson propagator. Similar to Eq. (8), we also have \cite{Qua_95,Sch_95} 
\begin{equation}		
g^{-4}_{\pi qq}=-N^2[I(0)+I(p)-m^2_{\pi}K(p)]^2,\quad \hbox{with}\quad (N=N_cN_f) \ . 
\end{equation} 
We note here that all above results are evaluated at threshold 
($p_i^2=p^2=m_{\pi}^2=s/4$), and we restrict them, for $T>0$, to the  
scattering of pions whose c.m. system is at rest in the heat bath\cite{Qua_95}. 
 
Summing Eqs. (5) and (6)and with the help of projection of the amplitudes on  
total isospin $I$, one finds the {\it s}-wave scattering amplitudes: 
\begin{eqnarray}		
{\cal T}^0 &=& 6{\cal T}_3+3{\cal T}_4+2{\cal T}_5 \nonumber \\ 
{\cal T}^1 &=& 0 \nonumber \\ 
{\cal T}^2 &=& 2{\cal T}_3+2{\cal T}_5 
\end{eqnarray} 
where the ${\cal T}_i$ are the function in Eqs.(5-6) excepted of the isospin 
factors. To get the density dependence on ${\cal T}_i$, we must  
calculate the constituent quark mass $m$ and the pion mass $m_{\pi}$  
at finite density first.  
 
As in Ref. \cite{Hat_87}, we get the in-medium constituent quark mass from  
the gap-equation due to the one-loop quark self-energy diagram 
(Fig. 5 in ref. \cite{Hat_87}). The  
in-medium pion mass (and also $\sigma$-meson mass) is determined from the  
dispersion relation (in the $\bfvec{q} \to \bfvec{0}$ limit) of the meson  
excitation (which itself is the solution of the Dyson equation in the  
ring-diagram approximation). For the details of deriving the gap-equation  
for quarks and the dispersion relation for $\pi$ (and also $\sigma$) meson,  
one can resort to Ref. \cite{Qua_95,Hat_87} with a straightforward extension
at finite density. Some specific results and integrals are explicitely given
in the appendix . Here we display only the final results   evolving with
respect to density at zero  temperature. One thing to be   mentioned here   is
that there are two different definitions for the coupling between pions  and
quarks, namely $g_{\pi qq}$: one is from Hatsuda {\it et al.}   \cite{Hat_87}
and the other is from Quack {\it et al.}  \cite{Qua_95}. These two  
$g_{\pi qq}$ differ from each other only of a term proportional to 
$$\beta\int{\rm d}k\frac{k^2}{E^2}f(E)f(-E)F(E,p)$$ 
where $\beta$ is the inverse of the temperature, $f(E)$ the  
Fermi-distribution function and $F(E,p)$ a rational fraction of 
$E$ ($E=\sqrt{m^2+\bfvec{k}^2}$) and $p$ (external momentum of 
pions). It turns out that  this term  disappears 
when the temperature goes to zero. So, the discrepancy between these two 
definitions will not give rise to any difference in the parameter fitting  
which is done only at zero temperature and also in the main part of this  
paper as we focus our calculation on the cases of finite density but at  
zero temperature.  
 
The density dependence of constituent quark mass, $\pi$ mass, $\sigma$  
mass and $f_{\pi}$ is shown in Fig. \ref{MasFp}. We find that all these  
quantities evolve with density almost linearly except in the high density  
region (say $\rho > 0.30 \ {\rm fm}^{-3}$) and the trends of these curves  
are all the same as those evolving with temperature \cite{Hat_87}. 
Here we use two sets of parameters listed in Tab. \ref{Paramt}.  
 
We first show that the density dependences of the threshold amplitudes  
and scattering lengths 
are very similar to those obtained at finite temperature in \cite{Qua_95}. 
In Fig. \ref{ScatAmp}, we draw the curves of  the various pieces of the  
scattering amplitudes as a 
function of density (at zero-temperature) with both sets of parameters.  
Again, we see that all the curves evolve linearly with density in the low and 
medium density region. At a certain density $\rho_d$, there exists a divergence  
as at high temperature \cite{Qua_95}.  The similar behavior also holds in 
Fig. \ref{ScatLen} where the relation between scattering length and  
density is illustrated. An interesting feature is that one gets a  
higher value of  density $\rho_d$ with the higher input vacuum 
$\sigma$-meson mass. This 
divergence occurring in the sigma propagator \cite{Qua_95}
actually corresponds to the point where the in-medium sigma mass drops to twice 
the in-medium pion mass.  
In the second set of parameters the density $\rho_d$ 
turns out to be $\rho_d\sim 2.4\rho_0$. Once unitarization is done (see section 5) 
this density goes down to about $\rho_d\sim 2\rho_0$ due to 
 the presence of the two-pion loop in the sigma propagator. 
We will come to this important point in section 5 since, as 
pointed out by Hatsuda {\it et al.} 
\cite{Hat_99}, the vanishing of the real part of the inverse 
sigma propagator at $E=2 m_\pi$ 
can be seen as a precursor effect of chiral symmetry restoration. 
However the aim of this paper is not primarily  to study the phase 
transition region  but the moderate density region below nuclear matter density
accessible in the $\pi-2\pi$ experiment. For that purpose, we show in 
Fig. 5 the low density behaviour of the scattering lengths.   
 
\section{The Simplified Scheme \hfill\ } 
Now we turn to the simplified scheme of our calculation, where we neglect 
the momentum dependence in all of above integrals, i.e. $I(p)$, $K(p)$ and 
$L(p)$ (we will state in the later of this section the reason for such a  
treatment). But before this simplified processing, we would like to check 
whether the Weinberg limit is  still  kept at finite density and finite  
temperature. First we want to see  
how these integrals (and then other related quantities) behave in the 
first order chiral expansion (expanding to the order of $m_{\pi}^2$). 
Keeping in mind that we are always in the 3-dimensional cut-off constraint 
and following the regularization of Ref. \cite{Sch_95}, we find after 
straightforward calculation, that 
\begin{eqnarray}		 
I(p^2=m_{\pi}^2) &=& I(0)-\frac{1}{3} m_{\pi}^2 K(0) \ , \nonumber \\ 
K(p^2=m_{\pi}^2) &=& K(0)-\frac{1}{2} m_{\pi}^2 L(0) \ , \nonumber \\ 
L(p^2=m_{\pi}^2) &=& L(0)-\frac{5}{6} m_{\pi}^2 I_{32}(0)   
\end{eqnarray} 
where $I_{32}(0)$ is shown explicitly in the appendix. 
With these expansions, we get from Eqs. (5-9)~:
\begin{eqnarray}		
{\cal T}_1 &=& \frac{4i}{N [2I(0)-\frac{1}{3}m_{\pi}^2 K(0) 
                             -m_{\pi}^2(K(0)-\frac{1}{2}m_{\pi}^2L(0))]} 
                             			     \ , \nonumber \\ 
{\cal T}_2 &=& {\cal T}_1                 	     \ , \nonumber \\ 
{\cal T}_3 &=& \frac{8i[I(0)+m_{\pi}^4(L(0)-\frac{5}{6}m_{\pi}^2 I_{32}(0))/2 
                         -2m_{\pi}^2(K(0)-\frac{1}{3}m_{\pi}^2L(0))]}  
                    {N [2I(0)-\frac{1}{3}m_{\pi}^2 K(0) 
                        -m_{\pi}^2(K(0)-\frac{1}{2}m_{\pi}^2L(0))]}  
                                 		     \ , \nonumber \\ 
{\cal T}_4 &=& \frac{[-8m(I(0)-\frac{1}{3}m_{\pi}^2K(0))]^2} 
                    {[2I(0)-\frac{1}{3}m_{\pi}^2 K(0) 
                       -m_{\pi}^2(K(0)-\frac{1}{2}m_{\pi}^2L(0))]^2} 
                                 		         \nonumber \\ 
	   & & \!\! \times  
	       \frac{i}{2N[(4m_{\pi}^2-4m^2)(I(0)-\frac{4}{3}m_{\pi}^2K(0)) 
	                   -m_{\pi}^2(I(0)-\frac{1}{3}m_{\pi}^2K(0))]} 
                             			     \ , \nonumber \\ 
{\cal T}_5 &=& \frac{[-8m(I(0)-m_{\pi}^2(K(0)-\frac{1}{3}m_{\pi}^2L(0))]^2} 
                    {[2I(0)-\frac{1}{3}m_{\pi}^2 K(0) 
                      -m_{\pi}^2(K(0)-\frac{1}{2}m_{\pi}^2L(0))]^2} 
                                 		         \nonumber \\ 
     	   & & \!\! \times 
     	       \frac{i}{2N[-4m^2I(0)-m_{\pi}^2(I(0)-\frac{1}{3}m_{\pi}^2K(0))]} 
	       					     \ . 
\end{eqnarray} 
Keeping to the first order of $m_{\pi}^2$  with the help of Eqs. (3) and (10), 
one finds finally that 
\begin{eqnarray}		
a^0 &=& \frac{1}{32\pi m_{\pi}} \frac{7 m_{\pi}^2}{f_{\pi}^2} \ ,\nonumber \\ 
a^2 &=& \frac{1}{32\pi m_{\pi}} \frac{-2 m_{\pi}^2}{f_{\pi}^2}  
\end{eqnarray} 
where $f_{\pi}$ is defined as in Ref. \cite{Hat_87}. Note there that the  
function $I(0)$, $K(0)$, $L(0)$  
and $I_{32}(0)$ are the integrals at finite density and  
finite temperature. We thus recover the Weinberg limit (in the 1st order  
chiral expansion) at threshold for finite density and finite temperature. 
 
In the simplified calculation, we neglect the momentum dependence of all  
above integrals inside the formula for ${\cal T}_i$, i.e., $p^2=m_{\pi}^2=0$  
except that we keep $k^2$ in Eq. (8) unchanged as the momentum transfer 
in the two-pion interaction. Thus one gets  
\begin{eqnarray}		
{\cal T}_1 &=& {\cal T}_2={\cal T}_3=\frac{2i}{NI(0)} \nonumber \\ 
{\cal T}_4 &=& {\cal T}_5=\frac{8im^2}{NI(0)}\frac{1}{k^2-m_{\sigma}^2}  
\end{eqnarray} 
where $m_{\sigma}^2=4m^2+m_{\pi}^2$ in the simplified calculation,  
which can be easily verified from Ref. \cite{Hat_87}. Since 
$f_{\pi}^2=-2iNm^2I(0)$ in the simplified calculation, the above amplitudes can be
rewritten as~:
\begin{eqnarray}		
{\cal T}_1 &=& {\cal T}_2={\cal T}_3=\frac{i}{NI(0)}=-2\lambda \nonumber \\ 
{\cal T}_4 &=& {\cal T}_5=4\lambda^2 f_{\pi}^2 \frac{1}{k^2-m_{\sigma}^2} 
\end{eqnarray} 
with the $\sigma\pi\pi$ coupling constant $\lambda$ defined by 
$2\lambda f_{\pi}^2=m_{\sigma}^2-m_{\pi}^2$ as in the  
linear sigma model (see for instance sect. 4 of \cite{Aou_94}). 
In other words, this simplified scheme yields  
exactly the amplitudes of the linear sigma model but with  
$\lambda$, $f_{\pi}$, $m_{\sigma}$ and $m_{\pi}$  
being  functions of density (and temperature). The numerical  
results for the simplified calculation are also displayed in  
Fig. \ref{MasFp}, \ref{ScatAmp} and \ref{ScatLen} by dashed lines. We find  
that these two calculations have no much difference from each other except  
for the high density region. So, the simplified calculation is a rather  
good approximation to address in our region of interest 
the density dependence of $m$, $m_{\pi}$,  
$m_{\sigma}$, $f_{\pi}$ and even the scattering amplitude and scattering  
length.

\section{In-medium Unitarized T Matrix} 
 
We now incorporate the previous $\pi-\pi$ amplitudes considered as a 
irreducible potential  
into a Lippman-Schwinger (LS) equation to get the unitarized T matrix in the  
scalar-isoscalar channel. However, when solving the LS integral equation, the  
use of the ``exact'' scheme  
with its full momentum dependence turns out to be of hopeless complexity. 
For this reason we prefer to use the potential from the ``simplified  
scheme'' which has been shown to be very close to the observables  
(pion decay constant, masses, scattering lengths) and their density  
dependences calculated with the ``exact'' scheme.  
Hence, the potential will be simply the linear sigma potential. 
For simplicity we restrict ourselves to an interacting pion pair with  
zero total momentum.  
In the scalar-isoscalar channel, the potential reads with obvious notations~: 
\begin{equation} 
<{\bf k}, -{\bf k}|V(E)|{\bf k}',-{\bf k}'>= v(k) v(k') 
{ m^2_\sigma-m^2_\pi\over f^2_\pi}\left( 
3 {s-m^2_\pi\over s-m^2_\sigma}+{t-m^2_\pi\over t-m^2_\sigma}+ 
{u-m^2_\pi\over u-m^2_\sigma}\right)\label{POT} 
\end{equation} 
The prescription for the Mandelstam variables to be used in the LS equation 
are the ones used in section 4 of \cite{Aou_94}. $E=\sqrt{s}$ is the energy  
variable at which the unitarized T matrix is calculated. $t$ and $u$ 
are chosen according to their on-shell values~: 
\begin{equation} 
t=2 m^2_\pi-2\omega_k \omega_{k'}+2{\bf k}\cdot{\bf k}',\qquad 
u=2 m^2_\pi-2\omega_k \omega_{k'}-2{\bf k}\cdot{\bf k}'\label{MAND} 
\end{equation} 
The form factor $v(k)$, which is assumed to simulate the momentum dependence  
of the vertices, is to be fitted on experimental phase shifts and scattering 
lengths in the free case. The unitarized T matrix is the solution of the LS 
equation~: 
\begin{eqnarray} 
&<&{\bf k}, -{\bf k}|T(E)|{\bf k}',-{\bf k}'> =  
<{\bf k}, -{\bf k}|V(E)|{\bf k}',-{\bf k}'>  \nonumber\\ & + &{1\over 2}\, 
\int {d^3{\bf q}\over(2\pi)^3} 
<{\bf k}, -{\bf k}|V|{\bf q},-{\bf q}>  
G_{2\pi}(E,{\bf q})  
<{\bf q}, -{\bf q}|V|{\bf k}',-{\bf k}'>\label{LS} 
\end{eqnarray} 
where the two-pion propagator in vacuum is \cite{Aou_94}~: 
\begin{equation} 
G_{2\pi}(E,{\bf q})={1\over \omega_q}\,{1\over E^2-4\omega^2_q+i\eta} 
\label{GPIPI} 
\end{equation}

Ignoring the $t$ and $u$ dependence in the denominators, the potential 
(Eq. (\ref{POT})) reduces 
to a combination of separable potentials. Taking parameter set II  
($m_\sigma=1$ GeV) implemented with a one-parameter form factor  
$v(k)=1/(1+k^2/64 m^2_\pi)$, one exactly recovers the potential used  
in \cite{Aou_94}. The phase shifts are correctly reproduced up to  
$800$ MeV and the scattering length, $a_0=0.23 m_\pi^{-1}$, is also  
in agreement with experimental data.  
It can be checked that, with such a separable potential, the correction 
to the scattering length from the unitarization procedure, is of higher  
order in $m^2_\pi$, thus preserving the chiral symmetry result  
(Weinberg scattering length) in the chiral limit.

From the underlying NJL quark model, the above results for the T matrix, 
Eq. (\ref{POT}), (\ref{MAND}), (\ref{LS}), (\ref{GPIPI})  
can be generalized at finite density by simply replacing the pion  
mass, the sigma mass and the pion decay constant by their in-medium  
values. With parameter set II, the pion mass does not change very much. Starting from  
the vacuum value $m_\pi=139$ MeV, one finds at half nuclear matter  
density $m_\pi(\rho_0/ 2)=140.3$ MeV, and at saturation density 
$m_\pi(\rho_0)=143$ MeV. The variation of the pion mass is usually   
expressed in term of the s-wave pion optical potential or in term  
of an effective scattering length $\Delta m^2_\pi=2m_\pi V_{opt}= 
-4\pi \left(b_0\right)_{eff}\rho$. Here we find $ \left(b_0\right)_{eff}\approx 
-0.01\, m_\pi^{-1}$ which coincides with the experimental pion-nucleon 
scattering length. In reality the pion optical potential is  
more repulsive due to higher order rescattering effects but still  
giving very small modification to the pion mass.        
 Hence, this modification of the pion mass, which  
simulates s-wave pion-nucleus interaction is practically negligible.  
In other words, the model does not give unrealistically large  
shift in the pion mass.  
 
On the contrary, there is a sizeable dropping 
of the sigma mass of $110$ MeV at half nuclear matter density and  
$225$ MeV are normal density. Consequently,  
one finds a significant reshaping of the $\pi-\pi$ strength  
function, as we will see in the next section, even at $\rho=0.5 \rho_0$  
which is the typical density reached in $\pi-2\pi$ 
experiment \cite{Kre_98}.  We now come to study the dressing  
of the pion propagator by its p-wave coupling in matter.  
As already stated in the introduction, the direct use of the 
standard NJL model for this  
problem would give totally unrealistic results since there is no repulsion  
and short range correlation ($g'$ parameter). Hence, we turn to a conventional  
nuclear matter description of the p-wave self-interaction of the pion on top 
of the NJL approach for the basic pion-pion potential. The pion propagator 
has the form~:    
\begin{equation} 
D_\pi({\bf k},\omega)\,=\,\big[\omega^2-m_\pi^2-{\bf k}^2  
 - S_\pi({\bf k},\omega)]^{-1}   
\end{equation} 
where $S_\pi({\bf k},\omega)$ is the conventional  nuclear matter p-wave  
pion self-energy to be specified below. Notice that $m_\pi$ is the  
in-medium pion mass calculated in the NJL model. According to the previous  
discussion, the small s-wave pion self-energy is partially accounted for  
but this gives in practice  a negligible effect for the in-medium $\pi-\pi$ 
T matrix.  For this first work, mixing the effects from the underlying  
quark structure (modification of $m_{\sigma,\pi}, f_{\pi}$)  
with those from pion p-wave coupling, we use the simplest possible approach. 
Hence, in the spirit of Ref. \cite{Cha_91}, we assume that the pion self-energy 
is dominated by virtual $\Delta-h$ excitations in the domain of energy of  
interest, namely~: 
\begin{equation}  
S_\pi({\bf k},\omega)\,=\,k^2\,\tilde\Pi^0({\bf k},\omega)\,= 
\,k^2\,\Pi^0({\bf k},\omega)\,/\,\left(1-g'_{\Delta\Delta} 
\Pi^0({\bf k},\omega)\right) 
\end{equation} 
Here $g'_{\Delta\Delta} \simeq 0.5$ accounts for the short range screening of  
the $\Delta-h$ polarization bubble $\Pi^0$. The numerical calculations are  
performed in the framework of the two-level model already used  
in Ref. \cite{Cha_91}. The main approximation of this model is to neglect the  
Fermi motion of the nucleons in the $\Delta-h$ bubble. In addition  
we first neglect the width of the delta resonance. This last approximation  
will be relaxed below. The polarization bubble has the form~: 
\begin{equation} 
\Pi^0({\bf k},\omega)\,=\,  
{4\over 9}\left({f^*_{\pi N \Delta}\over m_\pi} 
\Gamma(k)\right)^2\,\rho\,\left({1\over \omega-\epsilon_{\Delta k}+i \eta} 
-{1\over \omega+\epsilon_{\Delta k}}\right) 
\end{equation} 
with $\epsilon_{\Delta k}=\sqrt{k^2+M_\Delta^2}-M_N$ and  
$\Gamma(k)$ is the $\pi NN$ form factor. It is a simple matter  
to show that the pion propagator takes the very simple form typical of a  
two-level model~: 
\begin{equation} 
D_\pi({\bf k},\omega; \rho)\,= {Z_1(k,\omega ;\rho)\over \omega^2 - 
\Omega_1^2( k,\omega; \rho) + i \eta}\, 
+\,{Z_2(k,\omega ;\rho)\over \omega^2- 
\Omega_2^2(k,\omega; \rho) + i\eta} 
\end{equation} 
where the eigenenergies $\Omega_1, \Omega_2$ and the strength factors  
$Z_1, Z_2$ ($Z_1+Z_2=1$) are very simple functions of the density  
which can be found in Ref. \cite{Cha_91}. As explained in details in this  
last reference, the pion strength function splits into two collective  
eigenmodes. Of special interest is the lower branch ($\Omega_1$),  
sometimes called the pionic branch responsible for the softening  
of the pion dispersion curve. To calculate the medium modified  
$\pi-\pi$ T matrix, we have to solve the LS equation (Eq. (\ref{LS})) 
with a modified two-pion propagator according to~:  
\begin{equation} 
G_{2\pi}({\bf k},E)=\int {idk_0\over 2\pi}\, D_\pi({\bf k},k_0)\,  
D_\pi(-{\bf k},E-k_0) 
\end{equation}  
In the two-level model, its explicit expression is~: 
\begin{equation}   
G_{2\pi}({\bf k},E)=\sum_{i,j=1}^2\,{\Omega_i(k)+\Omega_j(k) \over 
2 \Omega_i(k)\, \Omega_j(k)}\, {Z_i(k) Z_j(k) \over E^2-\, \left(\Omega_i(k)+\ 
\Omega_j(k)\right)^2+ i \eta} 
\end{equation} 
Obviously, in the absence of p-wave coupling, one recovers Eq. (\ref{GPIPI}). 
In a more realistic description the main features of the two-level  
model survive but the modes $\Omega_{1,2}$ acquire a width. This can be  
incorporated through the replacement in the pion propagator (Eq. (23))~:  
\begin{equation} 
\Omega_j(k)\,\to\,\Omega_j(k)\,+\,i \,{{\rm Im} \tilde \Pi^0 \big({\bf k}, 
\Omega_j(k)\big)\over 2\, \Omega_j(k)}  
\end{equation} 
where ${\rm Im} \tilde \Pi^0 $, calculated along the line $j$,  
takes into account the delta width, corrected from Pauli blocking,  
together with extra 2p-2h contributions which are not reducible  
to a delta width piece. Details are given in Ref. \cite{Cha_93}.  
The imaginary part of the two-pion propagator is obtained  
through a spectral representation~: 
\begin{equation} 
{\rm Im} \, G_{2\pi}({\bf k}, E) =-{1\over \pi}\,\int_0^E\,d\omega\,  
      {\rm Im} D_\pi ({\bf k},E)\, {\rm Im} D_\pi({\bf k}, E-\omega)  
\end{equation} 
And the real part is calculated with a dispersion relation.

\section{Results}

To clearly illustrate the basic physical mechanisms, 
we  present the results for the in medium pion-pion scattering matrix in the
simplified calculational scheme discussed in the previous sections. 
First, the treatment of collective $\pi
N\Delta$ configurations entering the two-pion propagator is done in the extended
two-level model summarized just above. In addition we use the leading order term
of the so-called $1/N$ expansion of the linear sigma model. In a forthcoming
paper we will incorporate the full linear sigma model potential together with a
full calculation of the in-medium two-pion propagator. However, we believe that
the conclusions concerning the relative weight of collective pionic modes 
and chiral symmetry restoration will remain valid.

\smallskip
\noindent    
The linear sigma model can be seen as a $O(N+1)$ model with $N=3$. It has been shown 
\cite{Aou_97,Sch_98} that, to  leading order in a $1/N$ expansion, one gets a consistent 
symmetry conserving approach fulfilling Ward identities and all chiral symmetry
constraints. The corresponding potential is obtained from Eq. (\ref{POT}) by simply keeping
the $s$ channel pole  term and dropping the $u$ and $t$ sigma propagators. 
In practice it has the great advantage of making the potential separable. According 
to \cite{Sch_98}, we take the potential as~:  
\begin{equation}
<{\bf k}, \,-{\bf k}|V(E)|{\bf k}',\,-{\bf k}'>=
6\,\,v(k)\,v(k')\,\lambda
{E^2-m^2_\pi \over E^2-m^2_\sigma}
\quad\hbox{with}\quad
\lambda= {m^2_\sigma-m^2_\pi\over2\, f^2_\pi}
\end{equation}
with a phenomenological form factor $v(k)=g/(1+k^2/q_d^2)^\alpha$. It is important to
note that the introduction of such a form factor does not destroy in any way the
chiral symmetry properties. Taking $g=0.9$, $\alpha=3$ and $q_d=1\, GeV$ one gets 
a reasonable fit to experimental phase shifts \cite{Sch_98}.  With such a separable 
potential one easily obtains the explicit form of the $T$ matrix~:
\begin{equation}    
<{\bf k}, \,-{\bf k}|T(E)|{\bf k}',\,-{\bf k}'>=
6 \,v(k)\,v(k')\,\lambda
{E^2-m^2_\pi \over E^2-m^2_\sigma
-3\,\lambda\,(E^2-m^2_\pi) \,\Sigma(E)}
\end{equation}
with $\Sigma(E)$ 
given by~:
\begin{equation}
\Sigma(E)=\int{d^3q\over (2\pi)^3}\,v^2(q) \,G_{2\pi}({\bf q}, E)
\end{equation}
On figure 6, we present the results of the calculation for the $\pi - \pi$ strength 
function
({\it i.e.} the imaginary part of the scattering amplitude) for zero density, half
nuclear matter density ($0.5 \rho_0$)and normal nuclear matter density ($\rho_0$). 
We limit ourselves to the low energy sector ($E<600 MeV$) since it corresponds to  
the domain of validity of the low energy effective theory and 
to the region experimentally probed for instance in the CHAOS experiment. On Fig. 6a
we keep the sigma mass, the pion mass and the pion 
decay constant at their vacuum values 
and incorporate the effects of p-wave $\pi N \Delta$ collective modes. We recover the
well-known  structure near the two-pion threshold 
originating from  the softening of the pion 
dispersion relation and extensively discussed in \cite{Nor_88,Cha_91,Aou_94}. 
In Fig. 6b, we disregard the effect of these pionic modes but simply replace in the
vacuum T matrix the sigma mass, the pion mass and the pion decay constant by their 
in-medium value. 
In this way we isolate the effects intimately related to chiral symmetry 
restoration. We see a spectacular enhancement of the strength with increasing density
which is roughly uniform in the energy domain considered. The origin of this effect is
mainly the dropping of the sigma mass. However, it is less pronounced in the density
domain typical of ordinary nuclei than in the work of Hatsuda {\it et al.} 
\cite{Hat_99} for at least two obvious reasons. First, we start with a vacuum 
sigma mass of 1 GeV  (used in our previous works to fit the phase shifts)
significantly larger than the values of $550$ MeV or $750$  MeV used in 
\cite{Hat_99}. Second, the $\sigma\pi\pi$ coupling constant
$\lambda=m^2_\sigma - m^2_\pi/2f^2_\pi$  slightly decreases at variance with
the work of Hatsuda {\it et al.} where it was density independent. 
We will come to a more detailed comparison of the two
works when discussing the sigma propagator itself. 
In fig. 6c, we include simultaneously the two mechanisms. We recover the typical low
energy structure associated to pionic collective modes but significantly 
reenforced by the effect of chiral symmetry restoration. 

To study the evolution with density of scalar-isoscalar modes, it is also very
interesting to study the sigma meson spectral function {\it i.e.} the imaginary
part of the sigma propagator. It is a  simple matter to obtain the explicit
form of the sigma meson propagator in this simple model \cite{Sch_98}~:
\begin{equation}
D_\sigma(E)={1\over E^2\,-\,  m^2_\sigma\,-\,S_\sigma(E)}
\end{equation}
with the sigma self-energy given by~:
\begin{equation}
S_\sigma(E)={6 f^2_\pi \lambda^2\,\Sigma(E)\over 1-3\lambda \Sigma(E)}
\end{equation}
The results of the calculation are shown on figure 7 for the three 
previous cases~: pionic collective modes only (Fig. 6a), chiral
symmetry restoration only (Fig. 6b) and both (Fig. 6c). It is also interesting
to compare our results with the calculation of Ref.\cite{Aou_99} where 
chiral symmetry restoration is implemented by only dropping the sigma mass, 
ignoring  the modification of the $\sigma\pi\pi$ coupling constant. Since we
find in this present work that $\lambda$ is only slightly
modified (about  20\%  lower at $\rho=\rho_0$) it is not surprising  that the
reinforcement of the low energy enhancement is qualitatively similar in these
two works at least at normal nuclear matter density.   

\smallskip

Until now we have considered the density regime accessible in experiments 
on ordinary nuclei such as the CHAOS experiments. It is however 
extremely interesting to go beyond, although the basics nuclear physics 
ingredients of the calculation  are less under control. 
We now follow for a while the arguments of \cite{Hat_99}. Let us consider 
the spectral function of the sigma meson~:
\begin{equation}
\rho_\sigma(E)=-{1\over \pi}\,Im D_\sigma(E)=-{1\over \pi}\,
{Im S_\sigma(E)\over (E^2 - m^2_\sigma  - Re S_\sigma(E))^2
\,+\,(Im S_\sigma(E))^2}
\end{equation}
where $m_\sigma$ is as usual the in-medium sigma meson mass obtained here from the 
NJL model. As in the work of Hatsuda {\it et al.} \cite{Hat_99}, this in-medium
sigma meson mass has an evolution which follows linearly at low density the scalar
condensate. However the precise law of evolution is different since it comes 
from a nucleonic tadpole diagram in \cite{Hat_99}. Near the two-pion 
threshold the phase space factor yields to  the well-known behavior~:
\begin{equation}          
Im D_\sigma \sim Im S_\sigma \sim (E^2-4 m^2_\pi)^{1/2}
\end{equation}
 Before chiral symmetry restoration with complete 
$\sigma-\pi$ degeneracy, it must exist a density at which 
$Re D^{-1}_\sigma(E=2 m_\pi)=0$ or 
$4m^2_\pi-m^2_\sigma-S(E=2m_\pi)=0$. At such a density $\rho_d$,
ignoring the effect of pionic collective modes, the spectral function behaves as~:
\begin{equation}
\rho_\sigma(E\simeq 2 m_\pi)= -{1\over \pi}\,{1\over Im S_\sigma(E \simeq 2 m_\pi)} 
\sim \theta(E-2 m_\pi)/(E^2-4 m^2_\pi)^{1/2}
\end{equation} 
This implies that there arises a mild integrable singularity just above the two-pion
threshold in the medium. With our set of parameters this precursor effect of chiral symmetry
restoration occurs at a density slightly larger than $2\rho_0$ at variance with 
\cite{Hat_99} where this density was found at $1.25 \rho_0$. In addition when collective
pionic modes are included, there is strength below two-pion  
threshold  due to the various sources of width ($\Delta-h$, $2p-2h$). 
As a consequence, the mathematical singularity disappears. However, a precursor effect 
still manifests itself through a strong enhancement of the near-threshold structure
created by the p-wave pionic collective modes. This feature is illustrated on 
Fig. 7 where the influence of chiral symmetry restoration itself is visible by
comparison of Fig. 7a and Fig. 7b.

\section{Conclusion}

In this paper we have examined medium effects on the unitarized pion-pion interaction in the
scalar-isoscalar channel and the sigma meson spectral function. 
The new feature of this work is the simultaneous treatment of chiral symmetry restoration 
and nuclear effect associated with the existence of collective pionic modes.  Starting from the
standard Nambu-Jona-Lasinio model generalized at finite density, we have obtained an in-medium 
pion-pion potential. In particular, we have shown that the NJL model regenerates the linear 
sigma model potential but with modified parameters ($m_\sigma$, $m_\pi$ and $f_\pi$). 
This modification of the parameters intimately related to chiral symmetry restoration 
strongly reenforces the $\pi\pi$ strength function in the threshold region.  
Since the NJL model is not adapted to the calculation of collective pionic modes, we have used 
the standard phenomenological nuclear physics approach. When both effects are incorporated, 
the low-energy $\pi\pi$ structure resulting from the existence of nuclear collective modes 
is significantly enhanced in particular through the dropping of the sigma mass. 
This feature is obviously of considerable importance for the understanding of the structure observed 
by the CHAOS collaboration in $\pi-2\pi$ experiments on various nuclei.

{\bf Acknowledgments~:} We are indebted to Z. Aouissat, J. Delorme, M. Ericson,
J. Marteau, P. Schuck and J. Wambach for stimulating and enlightening
discussions.

\vfill\eject

\appendix
\begin{center}
{\huge \appendixname}
\end{center}

We list below the various integrals appearing in the text~:
 
\begin{equation}		
I(p)=\int \frac{{\rm d}^4k}{(2\pi)^4} \frac{1}{[k^2-m^2][(k+p)^2-m^2]} \ . 
\end{equation} 
 
\begin{equation}		
K(p)=\int \frac{{\rm d}^4k}{(2\pi)^4} \frac{1}{[k^2-m^2]^2[(k+p)^2-m^2]} \ . 
\end{equation} 
 
\begin{equation}		
L(p)=\int \frac{{\rm d}^4k}{(2\pi)^4} \frac{1}{[k^2-m^2]^2[(k+p)^2-m^2]^2} \ . 
\end{equation} 
 
\begin{equation}		
I_{mn}(p)=\int \frac{{\rm d}^4k}{(2\pi)^4}  
               \frac{1}{[k^2-m^2]^m[(k+p)^2-m^2]^n} \ . 
\end{equation} 

The above equations are strictly valid at $T = 0$. The explicit expressions
at finite temperature are given in \cite{Qua_95} and can be straightforwardly 
extended at both finite density and temperature. Here we  simply quote 
a few relevant results for numerical calculation at finite density~:
\begin{eqnarray}
-iI(p_0) & = & \int _{k_F} ^{\Lambda} \frac{dk}{2\pi
^2}\frac{k^2}{E}\frac{-1}{p_0 ^2 -4E^2} \cr
-iK(p_0) & = & \int _{k_F} ^{\Lambda} \frac{dk}{2\pi
^2}\frac{-k^2}{4E^3}\left[\frac{1}{p_0^2-4E^2}-
\frac{8*E^2}{(p_0 ^2-4E^2)^2}\right] \cr
-iL(p_0) & = & \int _{k_F} ^{\Lambda} \frac{dk}{4\pi
^2}\frac{k^2}{4E^3 p_0 ^2}\left[\frac{1}{p_0^2-4E^2}-
\frac{12E^2}{(p_0 ^2-4E^2)^2}
-\frac{64E^4}{(p_0 ^2-4E^2)^3} \right] \cr
\end{eqnarray} 
and the gap equation is~:
\begin{eqnarray}
M & = & m + 4N_cN_f g \int _{k_F} ^{\Lambda} \frac{k^2 dk}{2\pi ^2}\frac{M}{E}
\end{eqnarray} 

\begin{table}			
\caption{Physical quantities at $\rho=0$ in NJL model for different 
	 parameters (I and II) and different schemes of calculations  
	 (Exact and Simplified).} 
\begin{tabular}{cccccccc} 
 &   (MeV)   &   (MeV)   &    (MeV)     &   (MeV)   & (MeV) & (MeV) & (fm$^2$)\\
 
 & $f_{\pi}$ & $m_{\pi}$ & $m_{\sigma}$ & $\Lambda$ &  $m$  & $m_0$ & $g$     \\
 
\hline 
I(Exact) & 93.0 & 139.0  &    700.0     &   620.84  & 347.49& 5.677 & 0.2247 \\ 
II(Exact) & 93.0 & 139.0 &   1000.0     &   573.31  & 498.85& 5.772 & 0.3303\\ 
\hline 
I(Simplified) & 93.0 & 139.0 & 700.0    &   624.25  & 343.03& 5.540 & 0.2207\\ 
II(Simplified) & 93.0 & 139.0 & 1000.0  &   573.60  & 495.15& 5.712 & 0.3284 
\end{tabular} 
\label{Paramt} 
\end{table} 
 
\vfill\eject

\begin{figure}          	
\caption{Box and $\sigma$-propagation diagrams for $\pi$-$\pi$ 
	 scattering ($a,b,c$ and $d$ are the isospin indices.} 
\label{ScatFD} 
\end{figure} 
 
\begin{figure}          	
\caption{Constituent quark mass $m$, pion mass $m_{\pi}$,  
	 $\sigma$-meson mass $m_{\sigma}$ and pion decay constant $f_{\pi}$ 
	 vs. density (at zero-temperature) with parameter set I and II: solid 
	 lines for the exact calculation and dashed lines for the simplified 
	 calculation.} 
\label{MasFp} 
\end{figure} 
 
\begin{figure}          	
\caption{Scattering amplitudes ${\cal T}_i$ vs. density (at  
	 zero-temperature) with parameter set I and II: solid 
	 lines for the exact calculation and dashed lines for the simplified 
	 calculation.} 
\label{ScatAmp} 
\end{figure} 
 
\begin{figure}          	
\caption{Scattering lengths $a^0$ and $a^2$ vs. density (at  
	 zero-temperature) with parameter set I and II: solid 
	 lines for the exact calculation and dashed lines for the simplified 
	 calculation.} 
\label{ScatLen} 
\end{figure} 

\begin{figure}          	
\caption{Same as Fig. 4 but for moderate range of density. \protect \hfill} 
\label{MODER} 
\end{figure} 

\begin{figure}          	
\caption{Strength function for the $\pi - \pi$ interaction at zero density (dot-dashed curves), 
half nuclear matter density (dashed curves) and normal nuclear matter density (solid curves).
Fig. 6a (upper left panel) corresponds to a calculation 
with coupling to the pionic collective modes,  
keeping $m_\sigma$, $m_\pi$ and $f_\pi$ at their vacuum values. 
Fig. 6b (upper right panel) incorporates the
in-medium modification of $m_\sigma$, $m_\pi$ and $f_\pi$ from the NJL model, ignoring the
effect of collective pionic modes. Fig. 6c (lower panel) incorporates both
effects.} 
\label{IMT} 
\end{figure} 

\vfill\eject

\begin{figure}                  
\caption{Spectral function for the sigma meson at various densities. Fig. 7a
(upper panel) corresponds to a calculation with coupling to the pionic
collective modes, keeping $m_\sigma$, $m_\pi$ and $f_\pi$ at their vacuum
values. Fig. 7b (lower panel) incorporates the in-medium modification of
$m_\sigma$, $m_\pi$ and $f_\pi$ from the NJL model on top of collective pionic
modes.} 
\label{IMSIG} 
\end{figure} 

\vfill\eject

\begin{figure}                  
\begin{center}
  \begin{minipage}[b]{\linewidth}
    \centering\epsfig{figure=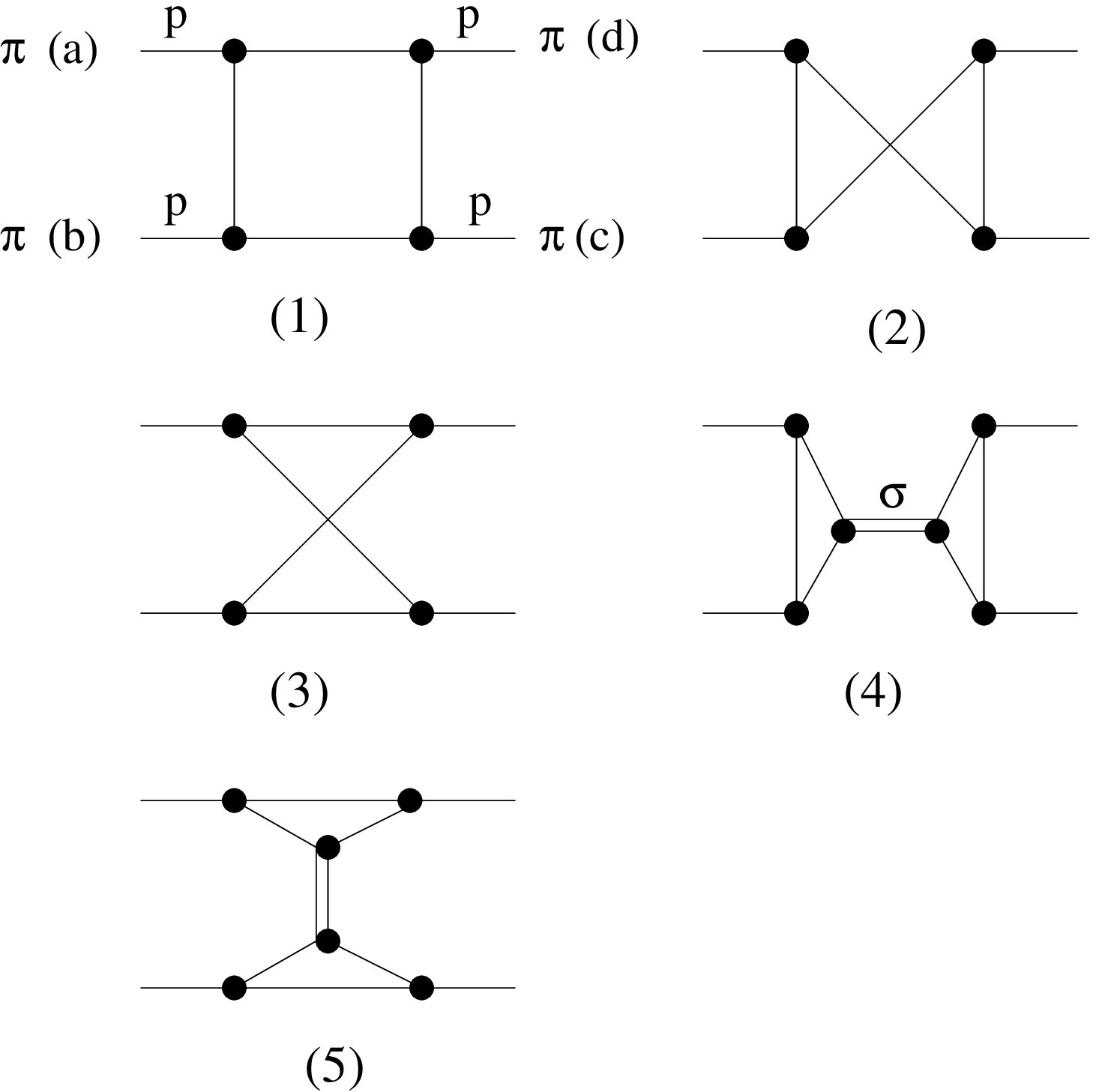,width=\linewidth}
\end{minipage}
\end{center}
\end{figure}

\vfill\eject

\begin{figure}                  
\begin{center}
  \begin{minipage}[b]{\linewidth}
    \centering\epsfig{figure=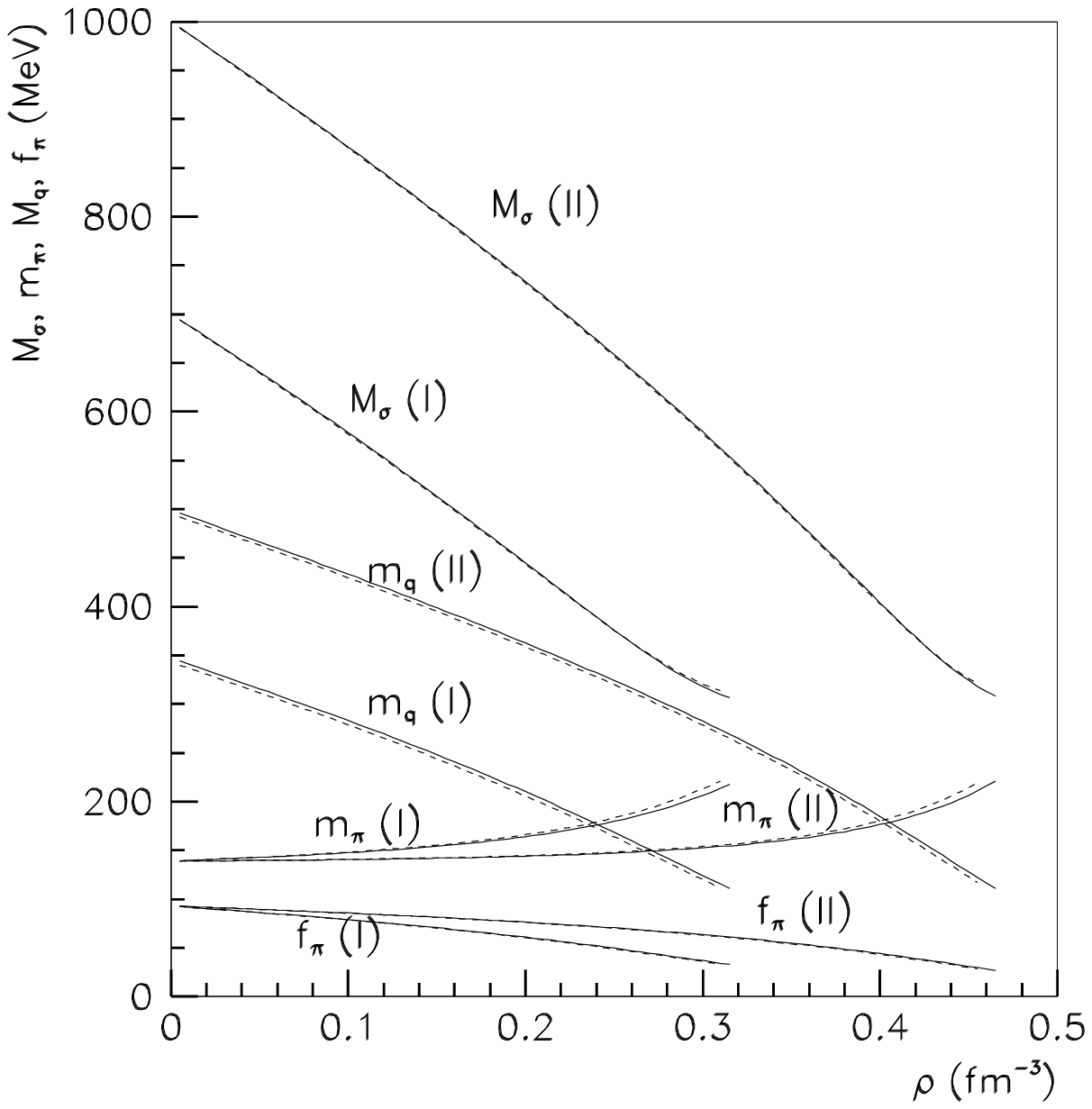,width=\linewidth}
\end{minipage}
\end{center}
\end{figure}

\vfill\eject

\begin{figure}                  
\begin{center}
  \begin{minipage}[b]{\linewidth}
    \centering\epsfig{figure=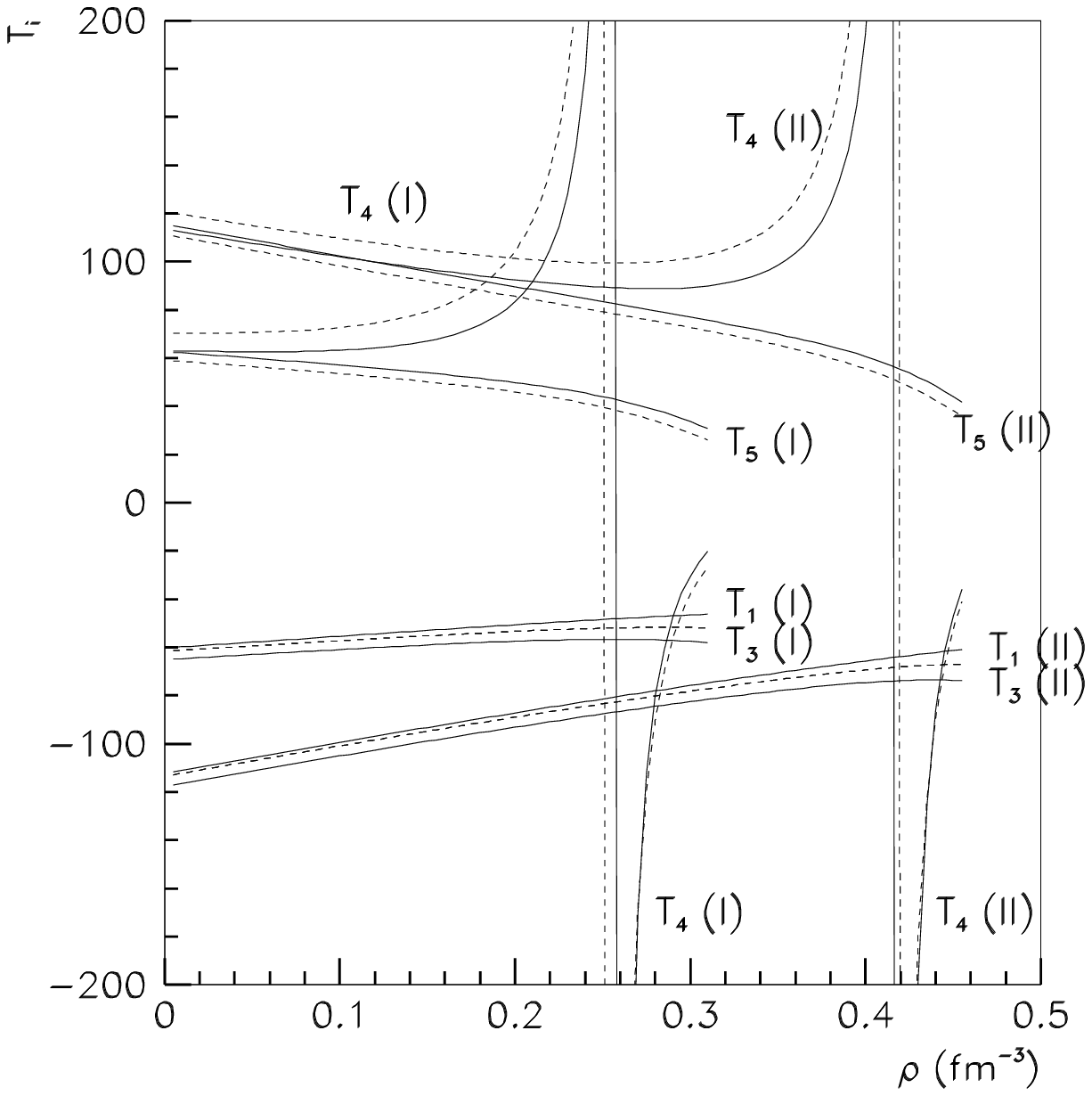,width=\linewidth}
\end{minipage}
\end{center}
\end{figure}

\vfill\eject

\begin{figure}                  
\begin{center}
  \begin{minipage}[b]{\linewidth}
    \centering\epsfig{figure=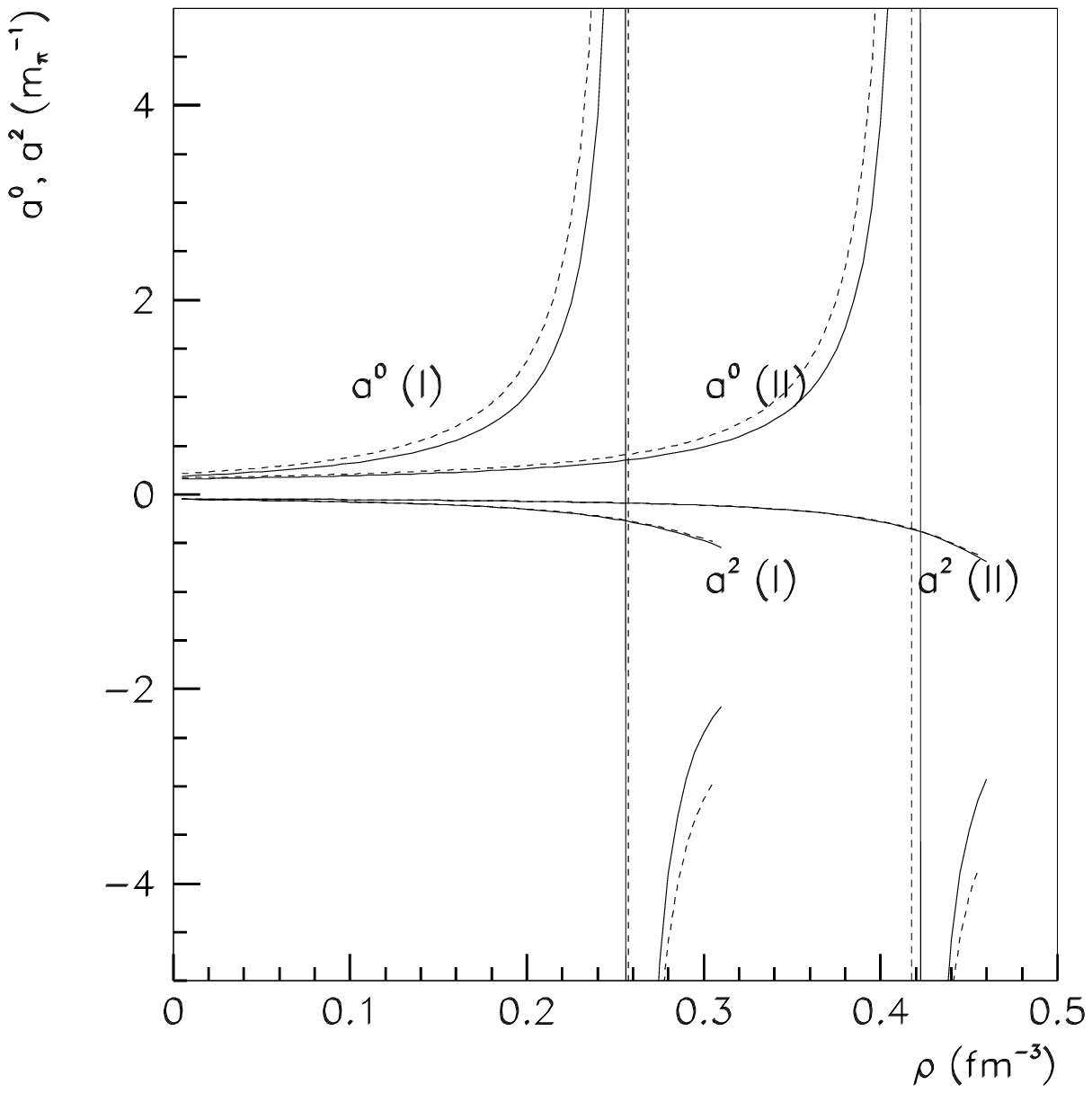,width=\linewidth}
\end{minipage}
\end{center}
\end{figure}

\vfill\eject

\begin{figure}                  
\begin{center}
  \begin{minipage}[b]{\linewidth}
    \centering\epsfig{figure=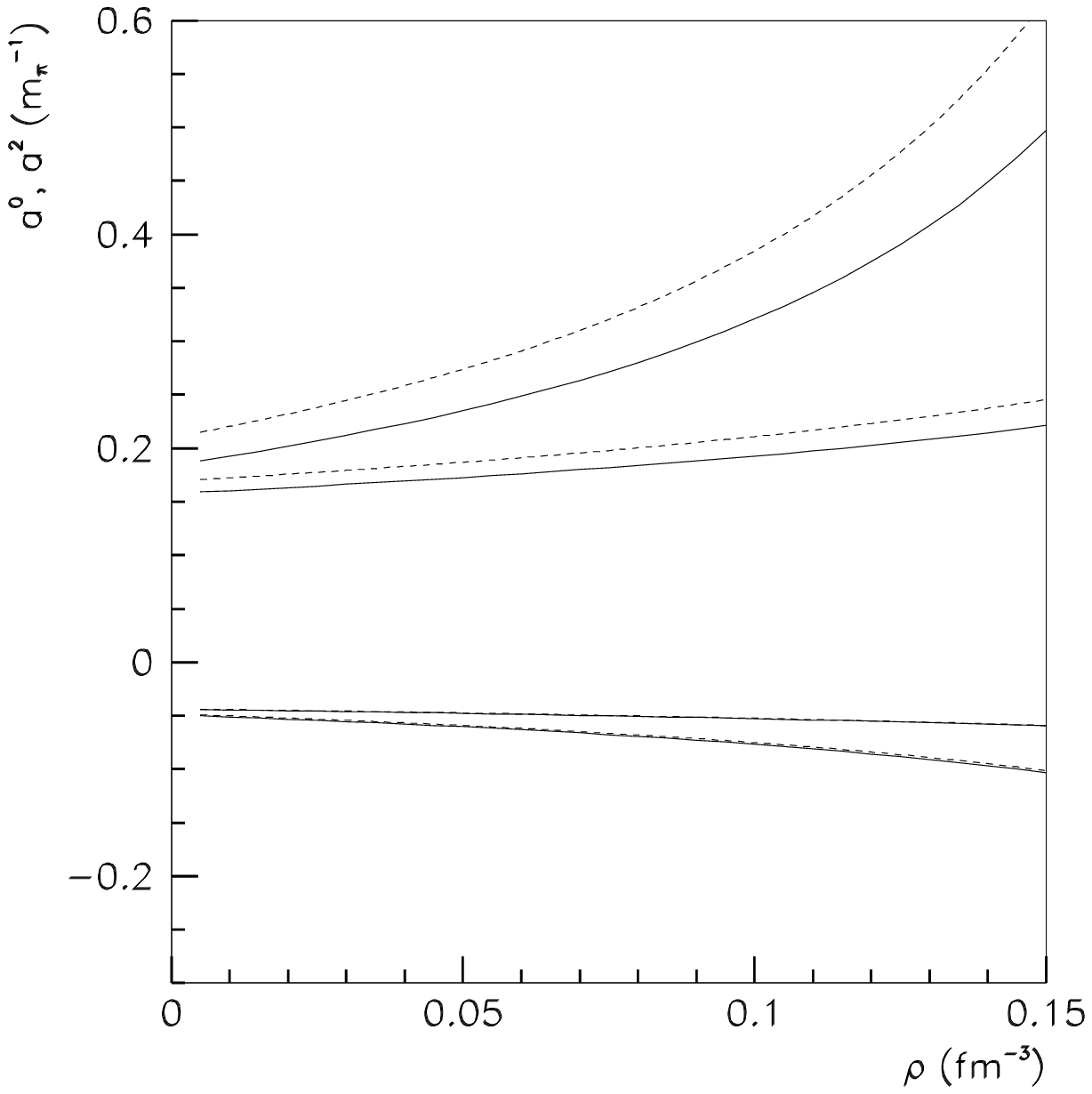,width=\linewidth}
\end{minipage}
\end{center}
\end{figure}

\vfill\eject

\begin{figure}                  
\begin{center}
  \begin{minipage}[b]{\linewidth}
    \centering\epsfig{figure=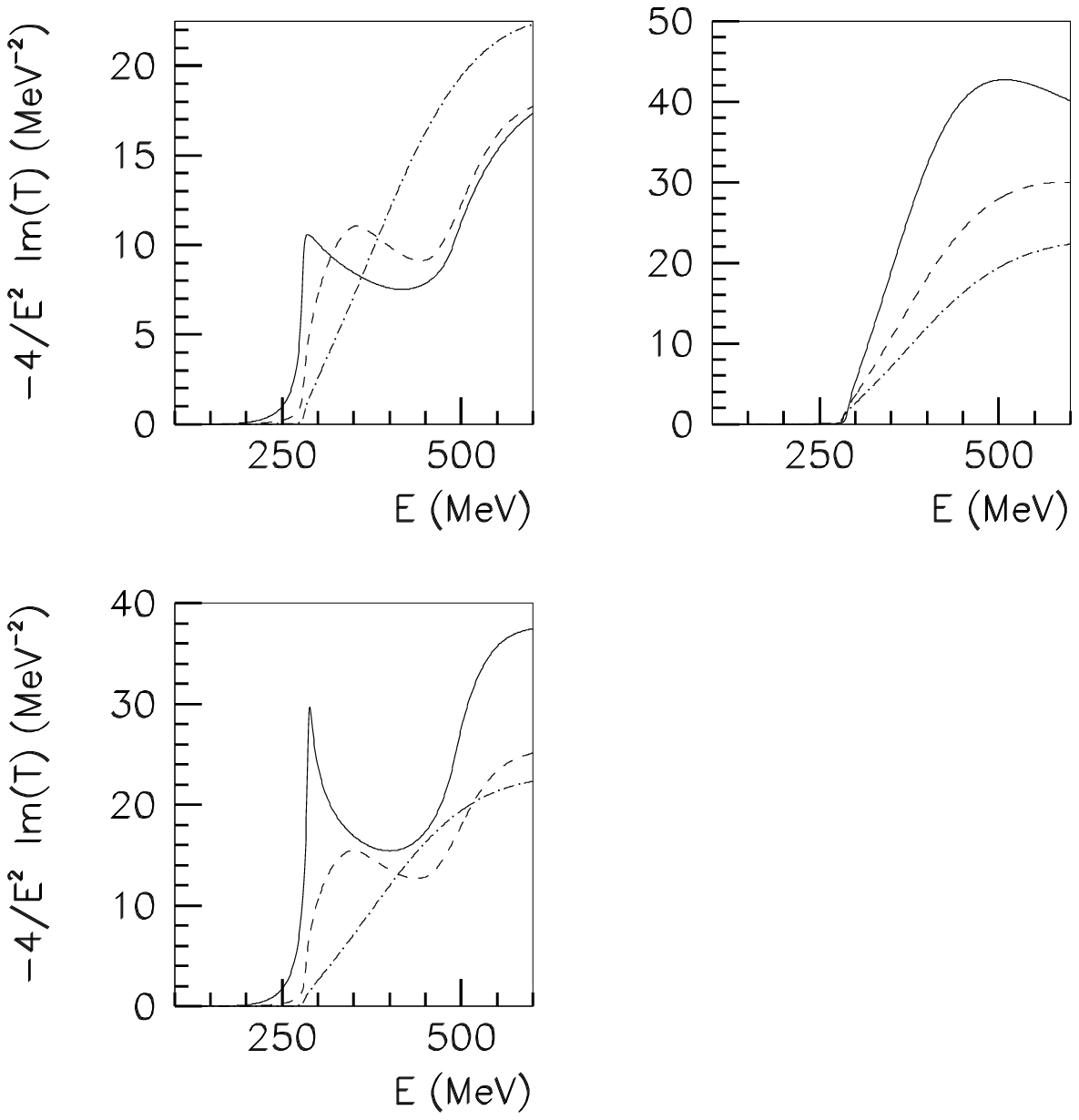,width=\linewidth}
\end{minipage}
\end{center}
\end{figure}

\vfill\eject

\begin{figure}                  
\begin{center}
  \begin{minipage}[b]{\linewidth}
    \centering\epsfig{figure=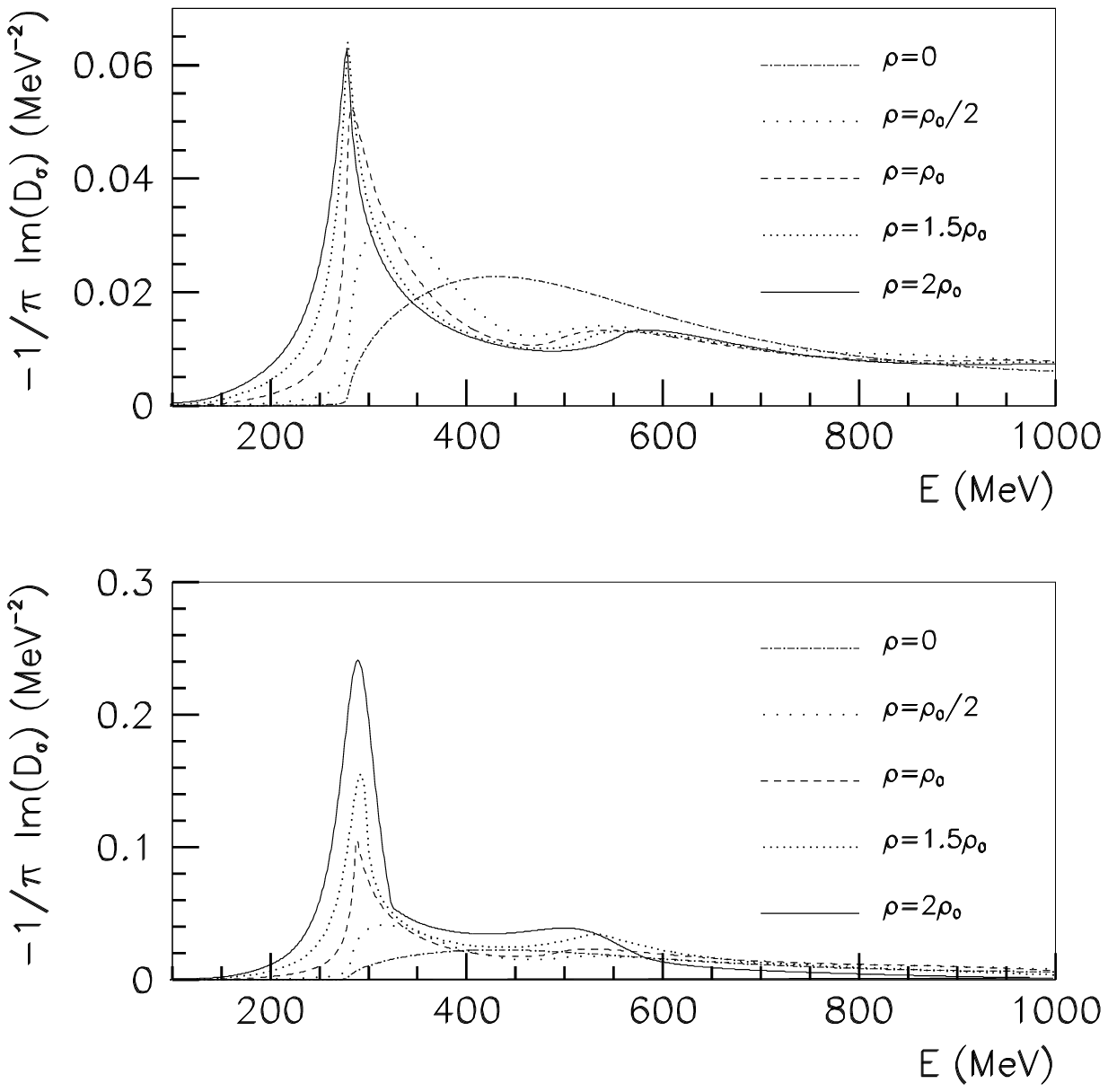,width=\linewidth}
\end{minipage}
\end{center}
\end{figure}

\vfill\eject 
 

\begin{thebibliography}{20} 
\bibitem{Nor_88} 
        P. Schuck, W. N\"{o}renberg and G. Chanfray, 
        Z. Phys. A 330 (1988) 119 
\bibitem{Gui_91} 
        P. Guichon and J. Delorme,  
	Phys. Lett B 263 (1991) 157 
\bibitem{Uda_90}  
        T. Udagawa, S. W. Hong and F. Osterfeld,  
	Phys. Lett B 245 (1990) 1 
\bibitem{Con_86}  
        D. Contardo {\it et. al.},  
	Phys. Lett. B 168 (1986) 331 
\bibitem{Hen_92}  
        T.Hennino {\it et. al.},  
	Phys. Lett. B 283 (1992) 42, 303 (1993) 236 
\bibitem{Cha_91} 
        G. Chanfray, Z. Aouissat, P. Schuck and W. N\"{o}renberg, 
        Phys. Lett. B 256, (1991) 325 
\bibitem{Aou_94} 
        Z. Aouissat, R. Rapp, G. Chanfray, P. Schuck and  J. Wambach,  
        Nucl. Phys. A 581 (1995) 471  
\bibitem{Dur_93}  
        J. W. Durso, H. C. Kim and J. Wambach,  
        Phys. Lett B 298 (1993) 267  
\bibitem{Rap_97}  
        R. Rapp, J. W. Durso and J. Wambach,  
        Nucl. Phys. A 615 (1997) 501 
\bibitem{Mac_97}  
        R. Rapp, R. Machleidt, J. W. Durso and G. E. Brown, 
        nucl-th/9706006  Phys. Rev. Lett. 82 (1999) 1827
\bibitem{Bon_96}  
        F. Bonutti {\it et al.} , the CHAOS collaboration, 
        Phys. Rev. Lett. 77 (1996) 603;  
\bibitem{Sch_98} 
        P.Schuck {\it et. al.},    
        nucl-th/9806069. 
\bibitem{Kre_98} 
        R. Rapp {\it et. al.}, 
        Phys. Rev. C 59 (1999) R1237
\bibitem{Vic99}
         M.J. Vicente Vacas and E. Oset,
         nucl-th/9907008 
\bibitem{Aou_99}
         Z. Aouissat, G. Chanfray, P. Schuck and  J. Wambach, 
         nucl-th/9908076 
\bibitem{Qua_95}		
	E. Quack, P. Zhuang, Y. Kalinovsky, S. P. Klevansky, and J. H\"{u}fner, 
	Phys. Lett. B 348 (1995) 1. 
\bibitem{Hat_99}
        T. Hatsuda, T. Kunihiro and H. Shimizu, 
        Phys. Rev. Lett. 82 (1999) 2840 
	T. Hatsuda and T. Kunihiro,
	nucl-th/9901020, nucl-th/9902025
\bibitem{Nam_61}         	
	Y. Nambu and G. Jona-Lasinio, 
	Phys. Rev. 122 (1961) 345; 124 (1961) 246. 
\bibitem{Hat_87}		
	T. Hatsuda and T. Kunihiro, 
	Prog. Theor. Phys. Suppl. 91 (1987) 284. 
\bibitem{Sch_95}		
	H. J. Schulze, 
	J. Phys. G 21 (1995) 185. 
\bibitem{Cha_93} 
        G. Chanfray and P. Schuck, 
        Nucl. Phys., A 555 (1993) 329 
\bibitem{Aou_97}
        Z.Aouissat, P. Schuck, J. Wambach,
        Nucl. Phys. A 618 (1997) 402  
\end{thebibliography}
\end{document}